# Predicting Surgery Duration from a New Perspective: Evaluation from a Database on Thoracic Surgery

Jin Wang, Ph.D.[a*], Javier Cabrera, Ph.D.[b], Kwok-Leung Tsui, Ph.D.[a], Hainan Guo, Ph.D.[c], Monique Bakker, Ph.D.[a], John B. Kostis, M.D.[d]

[a]Department of Systems Engineering and Engineering Management, City University of Hong Kong, Hong Kong, China; [b]Department of Statistics and Biostatistics, Rutgers University, NJ, USA; [c]Research Institute of Business Analytics and Supply Chain Management, College of Management, Shenzhen University, Shenzhen, China ; [d]Cardiovascular Institute, Rutgers Robert Wood Johnson Medical School, New Brunswick, NJ

**BACKGROUND**: Clinical factors influence surgery duration. This study also investigated non-clinical effects.

**METHODS**: 22 months of data about thoracic operations in a large hospital in China were reviewed. Linear and nonlinear regression models were used to predict the duration of the operations. Interactions among predictors were also considered.

**RESULTS**: Surgery duration decreased with the number of operations a surgeon performed in a day ($P < 0.001$). Also, it was found that surgery duration decreased with the number of operations allocated to an OR as long as there were no more than four surgeries per day in the OR ($P < 0.001$), but increased with the number of operations if it was more than four ($P < 0.01$). The duration of surgery was affected by its position in a sequence of surgeries performed by a surgeon. In addition, surgeons exhibited different patterns of the effects of surgery type for surgeries in different positions in the day.

**CONCLUSIONS**: Surgery duration was affected not only by clinical effects but also some non-clinical effects. Scheduling and allocation decisions significantly influenced surgery duration.

The duration of surgery depends on clinical or patient factors[1–4], e.g., age, gender, body mass index, American Society of Anesthesiologists risk class, etc. The characteristics of surgeons, such as surgeon's experience,[5,6], surgeon team size[7], and surgeon fatigue[8] also influence the duration of surgery.

Nonclinical factors have not been taken into consideration for predicting surgery durations. However, researchers have examined the effects of nonclinical factors on length of stay (LOS), mortality rate in hospital, and readmission rate. The typical nonclinical factors include the day of the week, the occupation level in the medical ward, and the physician's workload. It is shown that admission in weekends results into higher mortality rate[9,10]. Previous investigators have reported that higher workload results in the short LOS[11,12] and high readmission[12] and mortality rates[13]. High workload is found to be associated with distractions in operating rooms (OR)s[14]. The workload in (ORs) is different from that in medical wards where it is often represented by the number of patients in a given clinical unit, since the work is performed nearly simultaneously, while the operations of a surgeon are performed sequentially.

---

[*] Corresponding author. Tel.: +852 5981 4158
Email address: wangjin.devin@my.cityu.edu.hk

The aim of this study is to identify the factors that influence surgery duration from a new perspective. Specifically, this paper focuses on the effects of (1) the surgeon workload, (2) the workload in an OR, (3) the position in the sequence of surgeries performed by a surgeon, and (4) interactions of the factors on surgery durations.

## Methods

### Data source

Data used in this paper were obtained from the department of thoracic surgery in the First Affiliated Hospital of Dalian Medical University, Liaoning province, China. It is one of the largest hospitals in the province[15] and a top 100 hospital in China[16]. The department of thoracic surgery is the largest department in the hospital. The data from January 2014 through October 2016 included 2451 observations. 255 Observations with missing records (e.g., surgery type or the surgery start/end time was missing) were excluded. All operations included in this report were thoracic surgeries, of which pulmonary lobectomy accounted for more than half (67.39%) of the operations.

Compared with the ACS-NSQIP database that was often quoted in the literature (e.g., [1] and [4]), our database includes some additional variables, i.e., surgeon-specific variable, OR-specific variable and the variable on the detailed timeline and sequence of each surgery. Specifically, our data included the information about (1) the surgeon and anesthetist of each surgery, (2) the OR where a surgery was performed, and (3) the detailed timeline of each surgery (i.e., (i) the time of a patient entering the OR, (ii) the time of starting anesthesia, (iii) the time of the surgeon started surgery, (iv) the time of the surgeon finished the surgery, and (v) the time when patient left the OR). This allowed the analysis of the duration of surgery from the surgeon and OR perspectives, in addition to the analysis based only on patient factors. Note that the surgeons are anonymously presented throughout the paper to protect their privacy.

### Study variables

The primary variables included the surgeon performing the procedure, the number of surgeries a surgeon performed in a day, the position of the surgery in the sequence of surgeries performed by the surgeon in the day, the number of surgeries scheduled in the OR where the surgery was performed, and the day of the week when the surgery was performed. Other variables included the surgery type, the anesthetist. The hospital did not use the ICD-10 code. The surgeries were categorized to a specific surgery type according to an internal manual of the hospital. In our database, there were 11 surgery types.

### Outcome variables

Normally, the medical staff involved in an operation included nurses, anesthetists, a surgeon and assistants. The surgery process was divided into four parts according to the five time points mentioned above. The first two parts and the last part were mainly finished by nurses and anesthetists. The third part was mainly done by the surgeon and his or her assistants. Since this part is critical for the quality of surgery. We focused on effects of the surgeons, and the primary outcome in this paper was the length of the third part of the operations. The log of surgery duration was used as the outcome variable in order to avoid the skewed distributions.

## Statistical analysis

We first examined the effects of the primary variables. Since the two variables, the number of surgeries a surgeon performed in a day and the surgery position, are collinear, two linear regression models with either of the variables were used to examine the effects. The box-plot of the surgery duration in 2 different ranges (cut off less or equal to 4) of the number of surgeries in an OR showed a change point. Specifically, when there were 4 or fewer surgeries allocated in an OR, the surgery duration decreased with the number of surgeries in an OR, while the duration increased with the number if there were more than 4 surgeries in an OR. Hence, a piece-wise linear regression model was proposed.

To investigate whether there were interactions among the variables that influenced surgery duration, we built a regression tree, which provided evidence that there may be interactions among the three variables, surgery type, surgeon, and surgery position. We formulated the models with the three interactions, and examined the necessity of considering the interactions by comparing the new models with the original model, using analysis of variance. Since there were too many variables in the models with interaction terms, we performed variable selection by applying the LASSO method.

## Results and Discussion

### Surgery types

The regression results in Table 1 found that surgery duration greatly depended on the surgery type, as expected from the procedure difficulty. The results illustrated the differences of surgery duration for different surgery types. The baseline was the surgery with the most number of instances, named the thoracoscopic interior pulmonary lobectomy.

### Surgeons

The regression results in Table 1 illustrated that the mean of surgery duration was relevant to the surgeon performing the procedure. The mean of the duration of surgeries performed by Surgeon B ($P < 0.001$) and F ($P < 0.001$) was about 13 minutes and 34 minutes less than that performed by Surgeon A (the intercept is 4.97), while the duration of Surgeon C's surgeries ($P = 0.011$) was around 9 minutes longer than that of Surgeon A.

### The day of the week

Significant effects of the day of the week were observed (Table 1). The duration of surgeries on Tuesday ($P = 0.033$) and Friday ($P = 0.042$) was about 8 minutes longer than those on Saturday. Tuesday and Friday were the two busiest days while Sundays and Mondays had the fewest surgeries. Hence, longer duration on Tuesday might reveal a weekend effect, i.e., surgeons appeared to work at a slower pace when they went back to work after the weekend.

### The number of surgeries in an OR in a day

The results of Model I and II demonstrated that the change point of the workload in OR existed, as illustrated in Figure 1. When there were no more than four surgeries in an OR, surgery duration decreased around 8 minutes ($P < 0.001$) if one more surgery was allocated into the OR; when there were more than four surgeries in an OR, surgery duration increased around 12 minutes ($P < 0.01$) for one more surgery added to the OR. One possible reason was that if too many surgeries were allocated in an OR, the OR became disordered, which resulted in longer surgery duration.

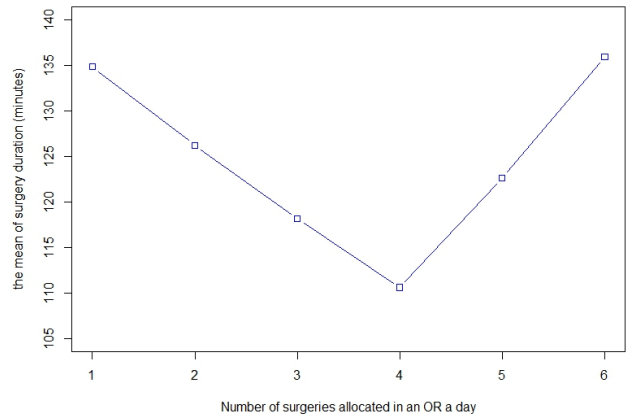

**Figure 1** The relationship between the mean of surgery duration and the number of the number of surgeries in an OR in a day.

### The number of surgeries performed by a surgeon in a day

The regression results of Model I showed that surgery duration decreased with the number of surgeries a surgeon performs in a day. That is, surgeons tended to accelerate their work if facing more surgeries. Numerically, the mean of surgery duration decreased by about 10 minutes ($P < 0.001$) if a surgeon performs one more surgery.

### The surgery position in the sequence in a day

As we mentioned before, the operations of a surgeon were performed sequentially. Hence, the workload pressure for surgeons may happen for the early surgeries, not for the later ones. The regression results of Model II verified this conjecture. Though some items were not significant when a surgeon performs five surgeries in a day, they only took up 1.18% of all cases. The coefficients were all negative, indicating that the mean of surgery duration decreased if a surgeon performed more than one surgery in a day. Also, the mean of the duration of the first surgery decreased more than later surgeries when a surgeon performed two, three or four surgeries in a day. For example, when a surgeon performed four surgeries in a day, the mean of the duration of the first surgery decreased by about 28 minutes ($P < 0.001$), while the mean of the duration of other surgeries decreased by about 20 minutes ($P < 0.05$).

Table 1 Regression results for models using only the main effects and without interaction terms. Anesthetist was excluded because of space.

| Perdictors | Frequency | Percentage | Results of Model I | | | | Results of Model 2 | | | |
|---|---|---|---|---|---|---|---|---|---|---|
| | | | Coefficient | 95% confidence interval | P value | | Coefficient | 95% confidence interval | P value | |
| **The day of the week**[†] | | | | | | | | | | |
| Sunday | 30 | 1.22% | -0.041 | (-0.188, 0.106) | 0.582 | | -0.045 | (-0.193, 0.104) | 0.556 | |
| Monday | 44 | 1.80% | -0.065 | (-0.188, 0.058) | 0.298 | | -0.077 | (-0.201, 0.048) | 0.227 | |
| Tuesday | 509 | 20.77% | 0.056 | (0.004, 0.109) | 0.036 | * | 0.057 | (0.005, 0.11) | 0.033 | * |
| Wednesday | 491 | 20.03% | 0.033 | (-0.02, 0.086) | 0.224 | | 0.035 | (-0.018, 0.088) | 0.197 | |
| Thursday | 435 | 17.75% | 0.007 | (-0.047, 0.062) | 0.790 | | 0.009 | (-0.046, 0.063) | 0.755 | |
| Friday | 563 | 22.97% | 0.054 | (0.003, 0.106) | 0.039 | * | 0.053 | (0.002, 0.105) | 0.042 | * |
| **Number of surgeries in the OR in a day**[#] | | | | | | | | | | |
| Less than or equal 4 | | | -0.066 | (-0.089, -0.043) | <0.001 | *** | -0.067 | (-0.09, -0.044) | <0.001 | *** |
| More than 4 | | | 0.103 | (0.034, 0.172) | 0.003 | ** | 0.109 | (0.04, 0.178) | 0.002 | ** |
| **Number of surgeries a surgeon performed in a day**[#] | | | -0.061 | (-0.08, -0.043) | <0.001 | *** | | | | |
| **Position**[‡] | | | | | | | | | | |
| 2 surgeries in a day | 977 | 39.86% | | | | | | | | |
| 2~1 | | | | | | | -0.075 | (-0.123, -0.027) | 0.002 | ** |
| 2~2 | | | | | | | -0.067 | (-0.115, -0.019) | 0.006 | ** |
| 3 surgeries in a day | 612 | 24.97% | | | | | | | | |
| 3~1 | | | | | | | -0.169 | (-0.234, -0.104) | <0.001 | *** |
| 3~2 | | | | | | | -0.145 | (-0.209, -0.081) | <0.001 | *** |
| 3~3 | | | | | | | -0.163 | (-0.228, -0.098) | <0.001 | *** |
| 4 surgeries in a day | 222 | 9.06% | | | | | | | | |
| 4~1 | | | | | | | -0.206 | (-0.317, -0.095) | <0.001 | *** |
| 4~2 | | | | | | | -0.141 | (-0.257, -0.026) | 0.017 | * |
| 4~3 | | | | | | | -0.129 | (-0.238, -0.02) | 0.021 | * |
| 4~4 | | | | | | | -0.148 | (-0.255, -0.042) | 0.006 | ** |
| 5 surgeries in a day | 29 | 1.18% | | | | | | | | |
| 5~1 | | | | | | | -0.287 | (-0.601, 0.027) | 0.073 | . |
| 5~2 | | | | | | | -0.051 | (-0.364, 0.262) | 0.749 | |
| 5~3 | | | | | | | -0.132 | (-0.445, 0.182) | 0.410 | |
| 5~4 | | | | | | | -0.443 | (-0.784, -0.102) | 0.011 | * |
| 5~5 | | | | | | | -0.105 | (-0.419, 0.208) | 0.510 | |
| **Surgeons**[§] | | | | | | | | | | |
| Surgeon B | 507 | 20.69% | -0.094 | (-0.08, -0.043) | <0.001 | *** | -0.098 | (-0.144, -0.052) | <0.001 | *** |
| **Surgeon C** | 484 | 19.75% | 0.058 | (-0.139, -0.048) | 0.011 | * | 0.059 | (0.013, 0.104) | 0.011 | * |
| **Surgeon D** | 296 | 12.08% | -0.006 | (0.013, 0.103) | 0.826 | | -0.008 | (-0.061, 0.045) | 0.767 | |
| **Surgeon E** | 170 | 6.94% | -0.034 | (-0.059, 0.047) | 0.329 | | -0.039 | (-0.109, 0.03) | 0.271 | |
| **Surgeon F** | 75 | 3.06% | -0.272 | (-0.104, 0.035) | <0.001 | *** | -0.270 | (-0.365, -0.176) | 0.000 | *** |
| **Surgery type△** | | | | | | | | | | |
| **1. Lung cancer** | 99 | 4.04% | 0.155 | (0.073, 0.237) | <0.001 | *** | 0.157 | (0.075, 0.239) | < 0.001 | *** |
| **2. Thoracoscopic pulmonary bullous resection** | 61 | 2.49% | -0.369 | (-0.476, -0.263) | <0.001 | *** | -0.375 | (-0.482, -0.267) | < 0.001 | *** |
| **3.Thoracoscopic partial pulmonary lobectomy** | 443 | 18.07% | 0.009 | (-0.037, 0.054) | 0.710 | | 0.009 | (-0.036, 0.055) | 0.684 | |
| **4.Total pneumonectomy** | 27 | 1.10% | 0.388 | (0.238, 0.538) | <0.001 | *** | 0.391 | (0.241, 0.541) | < 0.001 | *** |
| **5.Partial pulmonary lobectomy** | 280 | 11.42% | 0.110 | (0.056, 0.163) | <0.001 | *** | 0.108 | (0.054, 0.161) | < 0.001 | *** |
| **6.Thoracoscopic exploration** | 32 | 1.31% | 0.031 | (-0.106, 0.168) | 0.655 | | 0.033 | (-0.104, 0.169) | 0.641 | |
| **7.Pulmonary wedge resection** | 58 | 2.37% | -0.077 | (-0.182, 0.027) | 0.145 | | -0.077 | (-0.181, 0.027) | 0.147 | |
| **8.Esophageal cancer** | 50 | 2.04% | 0.774 | (0.661, 0.887) | <0.001 | *** | 0.773 | (0.660, 0.885) | <0.001 | *** |
| **9.Mediastinal tumor resection** | 142 | 5.79% | -0.170 | (-0.239, -0.101) | <0.001 | *** | -0.168 | (-0.237, -0.099) | 0.000 | *** |
| **10.Pulmonary tumor resection** | 329 | 13.42% | -0.043 | (-0.093, 0.007) | 0.091 | . | -0.043 | (-0.093, 0.007) | 0.089 | . |

Significant codes: 0 `***'; 0.001 `**'; 0.01 `*'; 0.05 `.'; 0.1 ` '.
† categorical variable, the baseline is Saturday.
# continuous variables.
‡ categorical variable, the baseline is the fact that a surgery is the only one surgery a surgeon performed in a day. The notation "2~1" means the first surgery in the day with two surgeries performed by a surgeon in the day.
§ categorical variable, the baseline is Surgeon A who performed the most surgeries.
△ categorical variable, the baseline is a surgery type, named thoracoscopic interior pulmonary lobectomy.

## Interactions

The coefficients relevant to surgeons, surgery types and their interactions are shown in Table 2 of the Appendix. The coefficients of other predictors were not listed in the table, since they were quite similar to those of Model II. The mean duration of surgeries performed by Surgeon B was less than that of Surgeon A, the baseline ($P < 0.001$). Surgeon B needed more time to perform Surgery 2 ($P = 0.040$) and 10 ($P = 0.020$), but less time for Surgery 6 ($P = 0.05$). The mean of the duration of Surgeon C's surgeries was more than that of Surgeon A ($P = 0.045$). Surgeon C was slower when performing Surgery 8 ($P = 0.024$). The mean durations of Surgeon D and E's surgeries did not come out. However, Surgeon D spent less time on Surgery 4 ($P = 0.061$) and 8 ($P = 0.014$), but more time on Surgery 10 ($P = 0.025$). Surgeon E needed more time to perform Surgery 6 ($P = 0.001$), but less time to perform Surgery 1 ($P = 0.032$) and 2 ($P = 0.006$). Surgeon F was generally much faster than Surgeon A ($P < 0.001$), especially when he/she was performing Surgery 3 ($P = 0.062$). Alternatively, from a surgery perspective, it was illustrated that the mean of the duration of Surgery 1, 4, 5, and 8 is larger than that of the baseline, while that of Surgery 2, 7, 9, and 10 was smaller. In addition, Surgery 2 was hard for Surgeon B, but easy for Surgeon E and F; Surgery 10 was hard for Surgeon B, C, and D.

The coefficients of surgeons and surgery positions, and their interactions are shown in Table 3 of the Appendix. It was shown that Surgeon B and C were flexible; the duration of these two surgeons' surgeries was significantly influenced by the surgery positions. In particular, Surgeon B and C was much faster for the surgeries in position 4~3 ($P < 0.001$) and position 5~4 ($P = 0.005$) respectively. Surgeon C and D performed slowly for surgeries in position 3~1 ($P = 0.036$) and position 5~1 ($P = 0.018$) respectively. Surgeon E needed more time for the second ($P = 0.10$) and third surgeries ($P = 0.013$) when he/she had three surgeries in a day.

The results in Table 4 in Appendix found that surgery duration was affected by the interaction between surgery type and the position. In other words, although surgery duration greatly depended on surgery type, the effects of surgery type were different for surgeries in different positions. For example, the duration of Surgery 2 decreased if it was the last surgery in a day with three surgeries performed by a surgeon ($P = 0.001$). A similar interpretation can be done for all the interaction items. In particular, Surgery 3 was flexible, since its duration increased if it was placed in the position 3~1 ($P = 0.033$) and decreased a lot if it was in the position 5~4 ($P = 0.017$).

## Conclusion

In this paper, the effects on surgery duration were investigated from a new perspective. Instead of only focusing on patient factors, non-clinical factors were studied, especially those relevant to surgeons. It was found that surgery duration was influenced by surgeons' workload and workload in the OR where the surgery was performed. The duration decreased with surgeons' workload. However, it did not monotonically decrease with the workload in the OR. Specifically, surgery duration decreased with the number of surgeries in the OR in a day if it was not more than four, while it would increase with the number if it was beyond four. Also, the duration of a surgery was impacted by the position of the surgery in a sequence of surgeries a surgeon performs in a day. In addition, the interactions among surgeons, surgery types, and surgery positions also influenced surgery duration. The study had significant strengths including the large numbers of operations, the uniformity of the administrative procedures in the hospital, which benefited the analysis of the effects of nonclinical factors.

The principal limitation of this study is that the results are obtained based only on one database of a department of a hospital. The surgery types were also limited, which mainly are thoracic surgeries. Hence, more data should be used to verify whether the results in this paper holds for other surgeons and surgery types. Additionally, the fact that a surgeon works faster (slower) does not mean that he/she is better (worse) at the surgery. Hence, another limitation is that outcomes of the surgeries (e.g., long of stay, readmission rate, and mortality rate) are not mentioned in this paper. The effect of surgery duration on healthcare quality is worthy of attention.

## Acknowledgments

This work was supported in part by Research Grants Council (RGC) Theme-Based Research Scheme under Grant T32-102/14-N, and in part by the National Natural Science Foundation of China (NSFC) under Grant 71701132 and Grant 61603321.

# Appendix

**Tables for regression results with interaction**

**Table 2** The regression results relevant to surgeons, surgery types and their interaction

| Variables | Coefficient | 95% confidential interval | Pr(>|t|) | |
|---|---|---|---|---|
| **Surgeon** | | | | |
| Surgeon B | -0.115 | [-0.161, -0.070] | <0.001 | *** |
| Surgeon C | 0.045 | [0.001, 0.090] | 0.045 | * |
| Surgeon F | -0.214 | [-0.312, -0.116] | <0.001 | *** |
| **Surgery type** | | | | |
| SurgeryType 1 | 0.175 | [0.096, 0.254] | <0.001 | *** |
| SurgeryType 2 | -0.356 | [-0.510, -0.202] | <0.001 | *** |
| SurgeryType 4 | 0.429 | [0.275, 0.583] | <0.001 | *** |
| SurgeryType 5 | 0.103 | [0.052, 0.153] | <0.001 | *** |
| SurgeryType 7 | -0.084 | [-0.184, 0.016] | 0.098 | . |
| SurgeryType 8 | 0.822 | [0.684, 0.960] | <0.001 | *** |
| SurgeryType 9 | -0.165 | [-0.231, -0.099] | <0.001 | *** |
| SurgeryType 10 | -0.116 | [-0.176, -0.056] | <0.001 | *** |
| **Interactions** | | | | |
| Surgeon B :SurgeryType 2 | 0.251 | [0.011, 0.490] | 0.040 | * |
| Surgeon B :SurgeryType 3 | 0.089 | [-0.022, 0.201] | 0.115 | |
| Surgeon B :SurgeryType 6 | -0.477 | [-0.813, -0.142] | 0.005 | ** |
| Surgeon B :SurgeryType 10 | 0.159 | [0.025, 0.294] | 0.020 | * |
| Surgeon C :SurgeryType 8 | 0.157 | [-0.102, 0.796] | 0.010 | * |
| Surgeon C :SurgeryType 10 | 0.347 | [0.037, 0.276] | 0.130 | |
| Surgeon D :SurgeryType 4 | -0.432 | [-0.885, 0.021] | 0.061 | . |
| Surgeon D :SurgeryType 8 | -0.288 | [-0.518, -0.058] | 0.014 | * |
| Surgeon D :SurgeryType 10 | 0.150 | [0.019, 0.281] | 0.025 | * |
| Surgeon E :SurgeryType 1 | -0.414 | [-0.793, -0.035] | 0.032 | * |
| Surgeon E :SurgeryType 2 | -0.337 | [-0.579, -0.096] | 0.006 | ** |
| Surgeon E :SurgeryType 6 | 0.429 | [0.181, 0.677] | 0.001 | *** |
| Surgeon F :SurgeryType 2 | -0.362 | [-0.826, 0.102] | 0.127 | |
| Surgeon F :SurgeryType 3 | -0.266 | [-0.545, 0.014] | 0.062 | . |

**Table 3** The regression results relevant to surgeons, surgery positions and their interaction

| Variables | Coefficient | 95% confidential interval | Pr(>|t|) | |
|---|---|---|---|---|
| **Surgeon** | | | | |
| Surgeon C | 0.074 | [0.031, 0.118] | 0.001 | *** |
| Surgeon F | -0.269 | [-0.359, -0.180] | <0.001 | *** |
| **Surgery position** | | | | |
| Position 3~1 | -0.155 | [-0.217, -0.092] | <0.001 | *** |
| Position 3~3 | -0.085 | [-0.146, -0.024] | 0.007 | ** |
| Position 4~1 | -0.165 | [-0.268, -0.061] | 0.002 | ** |
| Position 4~2 | -0.102 | [-0.211, 0.007] | 0.067 | . |
| Position 4~4 | -0.100 | [-0.200, -0.001] | 0.048 | * |
| Position 5~1 | -0.422 | [-0.755, -0.089] | 0.013 | * |
| **Interactions** | | | | |
| Surgeon B : Position 2~1 | -0.141 | [-0.218, -0.064] | <0.001 | *** |
| Surgeon B : Position 2~2 | -0.141 | [-0.221, -0.061] | 0.001 | *** |
| Surgeon B : Position 3~2 | -0.228 | [-0.366, -0.091] | 0.001 | ** |
| Surgeon B : Position 3~3 | -0.171 | [-0.319, -0.024] | 0.023 | * |
| Surgeon B : Position 4~3 | -0.434 | [-0.642, -0.225] | <0.001 | *** |
| Surgeon B : Position 5~5 | -0.393 | [-0.917, 0.131] | 0.141 | |
| Surgeon C : Position 3~1 | 0.150 | [0.010, 0.290] | 0.036 | * |
| Surgeon C : Position 3~2 | -0.258 | [-0.393, -0.123] | <0.001 | *** |
| Surgeon C : Position 5~4 | -0.756 | [-1.283, -0.229] | 0.005 | ** |
| Surgeon D : Position 5~1 | 0.980 | [0.169, 1.790] | 0.018 | * |
| Surgeon E : Position 3~2 | -0.361 | [-0.790, 0.068] | 0.099 | . |
| Surgeon E : Position 3~3 | -0.428 | [-0.766, -0.089] | 0.013 | * |

**Table 4** The regression results relevant to surgery position, surgery types and their interaction

| Variables | Coefficient | 95% confidential interval | Pr(>\|t\|) | |
|---|---|---|---|---|
| **Surgery position** | | | | |
| **Position 3~1** | -0.124 | [-0.189, -0.060] | <0.001 | *** |
| **Position 3~2** | -0.076 | [-0.131, -0.021] | 0.007 | ** |
| **Position 4~1** | -0.077 | [-0.184, 0.030] | 0.160 | |
| **Position 4~4** | -0.090 | [-0.190, 0.010] | 0.076 | . |
| **Surgery type** | | | | |
| **SurgeryType 1** | 0.165 | [0.088, 0.242] | <0.001 | *** |
| **SurgeryType 2** | -0.313 | [-0.415, -0.211] | <0.001 | *** |
| **SurgeryType 4** | 0.421 | [0.274, 0.569] | <0.001 | *** |
| **SurgeryType 5** | 0.126 | [0.076, 0.176] | <0.001 | *** |
| **SurgeryType 8** | 0.759 | [0.649, 0.868] | <0.001 | *** |
| **SurgeryType 9** | -0.161 | [-0.227, -0.096] | <0.001 | *** |
| **Interactions** | | | | |
| **SurgeryType 2 : Position 3~3** | -0.648 | [-0.227, -0.096] | 0.001 | *** |
| **SurgeryType 3 : Position 3~1** | 0.138 | [-1.033, -0.264] | 0.033 | * |
| **SurgeryType 3 : Position 5~4** | -0.637 | [0.011, 0.265] | 0.017 | * |
| **SurgeryType 4 : Position 3~2** | -0.824 | [-1.162, -0.112] | 0.032 | * |
| **SurgeryType 5 : Position 3~1** | -0.269 | [-1.579, -0.069] | 0.041 | * |
| **SurgeryType 5 : Position 4~1** | -0.489 | [-0.528, -0.010] | 0.030 | * |
| **SurgeryType 6 : Position 3~1** | 0.415 | [-0.931, -0.048] | 0.122 | |
| **SurgeryType 6 : Position 3~3** | -0.559 | [-0.111, 0.941] | 0.036 | * |
| **SurgeryType 6 : Position 4~4** | 0.546 | [-1.082, -0.037] | 0.150 | |
| **SurgeryType 7 : Position 3~3** | -0.408 | [-0.198, 1.291] | 0.031 | * |
| **SurgeryType 8 : Position 4~4** | 0.868 | [-0.778, -0.038] | 0.024 | * |
| **SurgeryType 9 : Position 4~1** | -0.418 | [0.115, 1.621] | 0.127 | |
| **SurgeryType 10 : Position 3~3** | -0.356 | [-0.955, 0.119] | <0.001 | *** |
| **SurgeryType 10 : Position 5~1** | -1.465 | [-0.506, -0.207] | <0.001 | *** |